\documentclass[fleqn,11pt,twoside]{article}

\usepackage{amsthm,amsthm,amssymb, color, xcolor,epsfig, graphics, subfigure, relsize, stix}

\usepackage{amsmath, graphicx, latexsym, lscape}

\usepackage{accents}

\DeclareMathOperator{\crl}{curl}

\DeclareMathOperator{\nos}{supp}
\DeclareMathOperator{\card}{card}
\DeclareMathOperator{\spam}{span}

\makeatletter
\newcommand{\copyrightnote}[2]{{\renewcommand{\thefootnote}{}
 \footnotetext{\small\it
\begin{flushleft}
 \copyright \ #1   #2  
\end{flushleft}}}}

\newcommand{\Name}[1]{\begin{flushleft}
                       \LARGE \bf #1
                       \end{flushleft}\vspace{-3mm}}

\newcommand{\Author}[1]{\begin{flushleft}
                       \it #1 \end{flushleft}}

\newcommand{\Address}[1]{\begin{flushleft}
                       \it #1 \end{flushleft}}

\newcommand{\Date}[1]{\begin{flushleft}
                      \small  \it #1 \end{flushleft}}

\newcommand{\evenhead}{Author \ name}
\newcommand{\oddhead}{Article \ name}

\renewcommand{\@evenhead}{
\hspace*{-3pt}\raisebox{-15pt}[\headheight][0pt]{\vbox{\hbox to \textwidth
{\thepage \hfil \evenhead}\vskip4pt \hrule}}}
\renewcommand{\@oddhead}{
\hspace*{-3pt}\raisebox{-15pt}[\headheight][0pt]{\vbox{\hbox to \textwidth
{\oddhead \hfil \thepage}\vskip4pt\hrule}}}
\renewcommand{\@evenfoot}{}
\renewcommand{\@oddfoot}{}

\long\def\@makecaption#1#2{%
  \vskip\abovecaptionskip
  \sbox\@tempboxa{\small \textbf{#1.}\ \ #2}%
  \ifdim \wd\@tempboxa >\hsize
    {\small \textbf{#1.}\ \ #2}\par
  \else
    \global \@minipagefalse
    \hb@xt@\hsize{\hfil\box\@tempboxa\hfil}%
  \fi
  \vskip\belowcaptionskip}

\newcommand{\JNMPnumberwithin}[3][\arabic]{%
  \@ifundefined{c@#2}{\@nocounterr{#2}}{%
    \@ifundefined{c@#3}{\@nocnterr{#3}}{%
      \@addtoreset{#2}{#3}%
      \@xp\xdef\csname the#2\endcsname{%
        \@xp\@nx\csname the#3\endcsname .\@nx#1{#2}}}}%
}

\newcommand{\resetfootnoterule} {
  \renewcommand\footnoterule{%
  \kern-3\p@
  \hrule\@width.4\columnwidth
  \kern2.6\p@}
}

\renewcommand{\footnoterule}{}

\makeatother

\theoremstyle{definition}
\theoremstyle{definition}
\newtheorem{tw}{Theorem}[section]
\newtheorem{defi}[tw]{Definition}

\newtheorem{prop}[tw]{Proposition}
\newtheorem{col}[tw]{Corollary}

\begin{document}

\renewcommand{\evenhead}{ {\LARGE\textcolor{blue!10!black!40!green}{{\sf \ \ \ ]ocnmp[}}}\strut\hfill 
Ł Chomienia and AM Grundland
}
\renewcommand{\oddhead}{ {\LARGE\textcolor{blue!10!black!40!green}{{\sf ]ocnmp[}}}\ \ \ \ \  
On Riemann wave superpositions obtained from the Euler system
}

\thispagestyle{empty}
\newcommand{\FistPageHead}[3]{
\begin{flushleft}
\raisebox{8mm}[0pt][0pt]
{\footnotesize \sf
\parbox{150mm}{{\textcolor{blue!10!black!40!green}{{\bf Open Communications in Nonlinear Mathematical Physics}}}
\ \ {Special Issue: Bluman}, 2025\\[0.1cm]
\strut\hfill 
ocnmp:16908,
pp #2\hfill {\sc #3}}}\vspace{-13mm}
\end{flushleft}}

\FistPageHead{1}{\pageref{firstpage}--\pageref{lastpage}}{ \ \ }

\strut\hfill

\strut\hfill

\copyrightnote{The authors. Distributed under a Creative Commons Attribution 4.0 International License}

\begin{center}

{\bf {\large A Special OCNMP Issue in Honour of George W Bluman}}
\end{center}

\smallskip

\Name{On Riemann wave superpositions obtained from the Euler system}

\Author{Łukasz Chomienia\textsuperscript{1}, Alfred Michel Grundland\textsuperscript{2,3}}

\Address{\textsuperscript{1}Department of Mathematics and Statistics, University of Jyv\"{a}skyl\"{a}, P.O. Box 35 (MaD), FI-40014, Jyv\"{a}skyl\"{a}, Finland\\

\textsuperscript{2}Centre de Recherches Math{\'e}matiques, Universit{\'e} de Montr{\'e}al, Succ. Centre-Ville, CP6128, Montr{\'e}al (QC) H3C 3J7, Canada
\\

\textsuperscript{3}D{\'e}partement de Math{\'e}matiques et d'Informatique, Universit{\'e} du Qu{\'e}bec à Trois-Rivières,
CP 500, Trois-Rivi{\'e}res (QC) G9A 5H7, Canada}

\Date{Received November 12, 2025; Accepted December 10, 2025}

\setcounter{equation}{0}

\smallskip

\noindent
{\bf Citation format for this Article:}\newline
Ł Chomienia and AM Grundland,
On Riemann wave superpositions obtained from the Euler system,
{\it Open Commun. Nonlinear Math. Phys.}, Special Issue:\,Bluman, ocnmp:16908, \pageref{firstpage}--\pageref{lastpage}, 2025.

\strut\hfill

\noindent
{\bf The permanent Digital Object Identifier (DOI) for this Article:}\newline
{\it 10.46298/ocnmp.16908  }
\strut\hfill

\begin{abstract}

\noindent 
The paper contains an analysis of the conditions for the existence of elastic versus non-elastic wave superpositions governed by the Euler system in (1+1)-dimensions. A review of recently obtained results is presented, including the introduction of the notion of quasi-rectifiability of vector fields and its application to both elastic and non-elastic wave superpositions. It is shown that the smallest real Lie algebra containing vector fields associated with the waves admitted by the Euler system is isomorphic to an infinite-dimensional Lie algebra which is the semi-direct sum of an Abelian ideal and the three-dimensional real Lie algebra. The maximal Lie module corresponding to the Euler system can be transformed, by an angle preserving transformation, to this algebra which is quasi-rectifiable and describes the behavior of wave superpositions. Based on these facts, we are able to find a parametrization of the region of non-elastic wave superpositions which allows for the construction of the reduced form of the Euler system.

\end{abstract}

\label{firstpage}


\section{Introduction}

This paper is dedicated to Professor George Bluman (The University of British Columbia), with whom the second author (AMG) has had several discussions on group analysis of differential equations over the course of several years.
	
In this paper, we present a summary of the recently obtained results concerning elastic and non-elastic wave superpositions admitted by the Euler system. For the sake of conciseness, we omit the proofs of the listed theorems, to be found in the relevant publications [4,5,6,7].
	
		Let us consider a first-order quasilinear homogeneous hyperbolic system
		\begin{align}\label{syst_PDEs}
			\begin{aligned}
				&\sum_{i=1}^{p}A^i(u)u_i=0,\quad u_i=\dfrac{\partial u}{\partial x^i},\quad i=1,...,p,\\ 
			&x=\left(x^1,...,x^p\right)\in\mathbb{R}^p,\quad u=\left(u^1(x),...,u^q(x)\right)\in\mathbb{R}^q,
            \end{aligned}
		\end{align}
		with the differential constraints
		\begin{equation}\label{deriv_u_decomp}
			\dfrac{\partial u^\alpha}{\partial x^i}=\sum_{s=1}^{k}\xi^s(x)\gamma_s^\alpha(u)\lambda_i^s(u),\quad\alpha=1,...,q,\quad i=1,...,p.
		\end{equation}
		The functions $(u^1,...,u^q,\xi^1,...,\xi^k)$ are considered to be the unknown functions of $x^1,...,x^p$. The vector fields $\left(\gamma_s,\lambda^s\right)$ satisfy the eigenvalue equations
		\begin{equation}\label{algebra-ic}
			\sum_{i=1}^{p}\sum_{\alpha=1}^{q}\left(A_\alpha^{i\beta}(u)\lambda_i^s\right)\gamma_s^\alpha=0,\quad\det(A^{i\beta}_\alpha\lambda_i^s)=0 ,\quad\beta=1,...,q,\quad\forall s=1,...,k.
		\end{equation}
		
\begin{tw}[Riemann wave solution {$[1,2]$}]
\ \\
Suppose that $(\lambda,\gamma)$ is a set of $C^1$ functions satisfying the algebraic equation \eqref{algebra-ic} and that $f:\mathbb{R} \to \mathbb{R}^q$ is an integral curve $\Gamma$ of the vector field $\gamma^{\alpha}(u)\frac{\partial}{\partial u^{\alpha}}$ on $\mathbb{R}^q$ with parameter $r$, i.e. 
\begin{equation}\label{hyperbolic_syst}
\Gamma:\hspace{3mm}\dfrac{df^\alpha}{dr}=\gamma^\alpha(f^1(r),...,f^q(r)),\quad \alpha=1,...,q.
\end{equation}
Then, the relations
\begin{equation}
    u^{\alpha} = f^{\alpha}(r),\quad r = \phi (\lambda_i(r)x^i)
\end{equation}
(where $\phi$ is an arbitrary function of one variable $\lambda_i(r)x^i$) constitute a solution of the system \eqref{syst_PDEs}, subjected to the constraints \eqref{deriv_u_decomp}, called a Riemann wave. The scalar function $r$ is called a Riemann invariant.
\end{tw}
In this paper, we discuss the process of interaction of two Riemann waves and introduce tools for constructing a simplified system, derived from (\ref{syst_PDEs}), allowing for the analysis of this interaction.

The involutivity conditions in the Cartan sense for the existence of Riemann $k$-wave solutions (resulting from the interactions of $k$ single Riemann waves) of the system (\ref{syst_PDEs}) under the conditions (\ref{deriv_u_decomp}), with the freedom of $k$ arbitrary functions of one variable, was established by Z. Peradzynski [3]. The necessary and sufficient conditions for the existence of these solutions require that the commutators of each pair of vector fields $\gamma_i$ and $\gamma_j$ be spanned by these fields

(i) \begin{equation}
[\gamma_i,\gamma_j]\in span \{\gamma_i,\gamma_j\},\quad i,j=1,...,k,\quad i\neq j
\end{equation}

and that the Lie derivatives of the one-forms $\lambda^i$ along the vector fields $\gamma_j$ be

(ii) \begin{equation}\label{conditions}
\mathcal{L}_{\gamma_j}\lambda^i\in span \{\lambda^i,\lambda^j\},\quad i,j=1,...,k,\quad i\neq j
\end{equation}

In what follows, we consider the (1+1)-dimensional case for which the conditions for the one-form $\lambda^i$ (\ref{conditions}) are identically satisfied. Thus, according to (\ref{hyperbolic_syst}), Riemann waves are associated only with the vector fields $\gamma_s$, $s=1,...,k$.
We look for the solutions parametrized in terms of Riemann invariants which require that all vector fields $\{\gamma_1,...,\gamma_k\}$ commute among themselves. To ensure that this takes place, it is usually necessary that these vector fields be rescaled through certain functions $h_1,...,h_k$ in such a way that
\begin{equation}\label{commut_hgamma}
[h_i\gamma_i,h_j\gamma_j]=0,\quad i,j=1,...,k,\quad i\neq j.
\end{equation}
To achieve this result we make use of the following theorems.

\begin{tw}[Straightening of vector fields {$[4]$}]\label{recti_thm}
\ \\
Let $X_1,...,X_r$ be a family of vector fields defined on an $n$-dimensional manifold $N$ such that
		\begin{equation*}
			X_1\wedge...\wedge X_r\neq0\quad\text{at any point on }N.
		\end{equation*}
		There exists a local coordinate system $\left\{x^1,...,x^n\right\}$ on $N$ such that the first integrals of each vector field $X_i$ are given by the equations
		\begin{equation}\label{str8}
			x^1=k_1,...,x^{i-1}=k_{i-1},\quad x^{i+1}=k_{i+1},...,x^n=k_n
		\end{equation}
		for some constants $k_1,...,\hat{k}_i,...,k_n\in\mathbb{R}$, where $i$ denotes skipped element, if and only if the commutators for each pair of vector fields $X_i$ and $X_j$ are spanned by these fields
		\begin{equation}\label{Thm_2.1}
			\left[X_i,X_j\right]=f_{ij}^iX_i+f_{ij}^jX_j,\quad 1\leq i<j\leq r
		\end{equation}
		for a family of $r(r-1)$ functions $f_{ij}^i,f_{ij}^j\in C^\infty(N)$ with $1\leq i<j\leq r<n$.
\end{tw}

Note that in (\ref{Thm_2.1}) and in what follows, we do not assume the summation convention.

We now introduce the definition distinguishing the family of vector fields which satisfies (\ref{Thm_2.1}).

\begin{defi}[Quasi-rectifiability property {$[4]$}]\label{recti_def}
\ \\
\noindent 
A family of vector fields $X_1,...,X_r$ on $N$ is said to be quasi-rectifiable if there exists a local coordinate system $\left\{x^1,...,x^n\right\}$ on $N$ such that each vector field $X_i$ has the form
		\begin{align}\label{coord_syst}
				&X_i=g^i\left(x^1,...,x^n\right)\dfrac{\partial}{\partial x^{(i)}}\quad \text{ (no summation)},\quad i=1,...,r,\\
				&\prod_{i=1}^{r}g^i\left(x^1,...,x^n\right)\neq0\,\, \Rightarrow \,\,X_1\wedge \ldots \wedge X_r\neq 0\, \text{at any point on }N
		\end{align}
		for $g^1,...,g^r\ :\ N\rightarrow\mathbb{R}$. Otherwise the family $X_1,...,X_r$ is called non quasi-rectifiable.  
		
		The coordinate expression (\ref{coord_syst}) is called a {quasi-rectifiable form} for the vector fields $X_1,...,X_r$ and accordingly, their basis is also called quasi-rectifiable. We extend this term to Lie modules and algebras calling them quasi-rectifiable if they are spanned by vector fields satisfying condition (\ref{coord_syst}).
Note that the definition above is base dependent.
\end{defi}

We have shown [4] that quasi-rectifiability of vector fields $X_1,...,X_r$ ensures the existence of the proper rescaling functions $h_1,...,h_r$ in (\ref{commut_hgamma}). Namely, we have

\begin{tw}[The rescaling theorem ${[4]}$]\label{modified}
\ \\
\noindent 
Let $X_1,...,X_r$ be a family of vector fields defined on an $n$-dimensional manifold $N$ such that 
		\begin{equation*}
			(i)\qquad X_1\wedge...\wedge X_r\neq0\quad\text{at any point on }N\hspace{5mm}
			\end{equation*}
			 and
			 \begin{equation*}
			(ii)\qquad \left[X_i,X_j\right]=f_{ij}^iX_i+f_{ij}^jX_j,\quad 1\leq i<j\leq r
		\end{equation*}
		for certain functions $f_{ij}^i,f_{ij}^j\in C^\infty(N)$. Then there exist nonvanishing functions\\ $h_1,...,h_r\in C^\infty(N)$ such that all rescaled vector fields $h_1X_1,...,h_rX_r$ commute between themselves, i.e.
		\begin{equation}\label{rescvf}
			\left[h_iX_i,h_jX_j\right]=0,\quad 1\leq i<j\leq r.
		\end{equation}
\end{tw}
\bigskip

In what follows, we determine the rescaling functions $h_i \in C^{\infty}(N),\; i\in \{1,...,r\}$.

\begin{tw}[Integrating factor ${[4]}$]\label{integfac}
\ \\
\noindent 
Let $X_1,...,X_r$ be a quasi-rectifiable family of vector fields on $N$ and let $\mathcal{D}$ be the distribution spanned by $X_1,...,X_r$. Let $\eta_1,...,\eta_r$ be dual one-forms to each of the vector fields $X_1,...,X_r$ on $N$, respectively, i.e.
			\begin{equation}
				\eta_i(X_j)=\delta_i^j,\quad i,j=1,...,r.
			\end{equation}
			The nonvanishing functions $h_1,...,h_r\in C^\infty(N)$ are such that $h_1^{-1} X_1,\ldots,h_r^{-1} X_r$ commute among themselves if and only if each of the one-forms $h_i\eta_i$, restricted to a leaf of the distribution $\mathcal{D}$, is an exact differential
			\begin{equation}\label{f_commute_if}
				d\left(h_i\eta_i\right)\Big|_{\mathcal{D}}=0,\quad i=1,...,r.
			\end{equation}
\end{tw}
The relation (\ref{f_commute_if}) allows us to determine the rescaling functions $h_1,...,h_r$ which satisfy (\ref{rescvf})
\begin{col} $[4]$
\ \\
\noindent 
If each one-form $h_i\eta_i,\; i\in \{1,...,r\}$ restricted to $\mathcal{D}$ satisfies the condition 
\begin{equation}
h_i\eta_i|_{\mathcal{D}} = dx^i|_{\mathcal{D}},\quad i\in \{1,...,r\},
\end{equation}
then the coordinate system $x^1,...,x^r$ on $\mathcal{D}$ satisfies the equations 
\begin{equation}
X_ix^j = 0 \quad \text{for}\quad i\neq j,\quad i,j \in \{1,...,r\}.
\end{equation}
Hence, each vector field $X_i$ has the quasi-rectifiable form (\ref{coord_syst}).
\end{col}
Note that if there exists a pair of vector fields $X_i,X_j$ which do not satisfy the assumptions of Theorem 1.4, then the family of vector fields $X_1,...,X_r$ is not quasi-rectifiable. 

Let us add that for the particular case of a family of three vector fields ($k=3$), the simplified criteria for quasi-rectifiability, equivalent to Definition 1.3, have been established [6].

\begin{tw}\label{curv} $[6]$
\ \\
\noindent
            Let $\mathcal{F}=\{\gamma_1,\gamma_2,\gamma_3\}$ be a family of vector fields defined in the space $\mathbb{R}^3$ satisfying $\gamma_1\wedge \gamma_2\wedge \gamma_3 \neq 0$ at any point on $\mathbb{R}^3.$ The family $\mathcal{F}$ is quasi-rectifiable if and only if for any point $p\in \mathbb{R}^3$ the following integral tends to zero,
            \begin{equation}\label{vanish}
            \lim_{r \to 0}\frac{1}{\pi r^2}\int_{S^1_r(p)}\gamma_i \times \gamma_j \cdot d\sigma_r=0\; \; \text{for every}\; \;  i,j \in \{1,2,3\},\quad i\neq j,
            \end{equation}
            where $S^1_r(p) \subset \mathcal{D}_p$ is the one-dimensional sphere and  $\mathcal{D}$ is a two-dimensional distribution spanned by the vector fields $\gamma_i$ and $\gamma_j.$
\end{tw}
\noindent Note that, when $r \to 0,$ then $\gamma_i \times \gamma_j$ tends to the normal vector at point $p.$\\
As a consequence of Theorem 2.4, we have

\begin{col}$[6]$
\ \\
    The family of linearly independent vector fields $\{\gamma_1,\gamma_2,\gamma_3\}$ in $\mathbb{R}^3$ is quasi-rectifiable if and only if for any $i,j\in \{1,2,3\},\; i\neq j,$ we have $\crl (\gamma_i\times \gamma_j) \in \spam\{\gamma_i,\gamma_j\}.$
\end{col}   

\section{Elastic and non-elastic wave superpositions}

Let us now consider the case of two interacting Riemann waves, i.e. when $k=2$ in (\ref{deriv_u_decomp}).

The set 
\begin{equation}
            M:=\Bigg\{(t,x)\in \mathbb{R}^2_+:\; \nos\xi^1(t,\cdot)\cap \nos\xi^2(t,\cdot)\neq \emptyset,\; t\in (t_{\min},t_{\max})\Bigg\}\subset\mathbb{R}^2
            \label{reggionM}
            \end{equation}
            is called the region of superposition of two waves corresponding to the vector fields $\gamma_1, \gamma_2$, where $(t,x)$ are time and space variables, respectively.
Let $M_{\epsilon}$ be an $\epsilon$-neighborhood of the region $M$ and $M_{2\epsilon}$ be a $2\epsilon$-neighborhood of the region $M_{\epsilon}$. We define the set 
            \begin{equation}\label{colar}
            A_{\epsilon}:= M_{2\epsilon} \setminus M_{\epsilon}.
            \end{equation}
            Let $B_{\delta} \subset \mathbb{R}^2_+$ be a ball of radius $\delta>0$ and $\phi_i^t$ the flow of the vector field $\gamma_i$. Then the set of two entering waves, according to (\ref{deriv_u_decomp}), is defined by
            $$
            \Gamma_+:= \{\gamma_i: du=\sum^2_{i=1}\xi^i\gamma_i\otimes \lambda_i,\; \exists B_{\delta}\subset u(A_{\epsilon}): \phi_i^t(B_{\delta})\subset u(M) \text{ for some } t \in (0,T)\}
            $$
            
            and the set of several leaving waves is defined by
            $$
            \Gamma_-:= \{\gamma_i: du=\sum^{k\geq 2}_{i=1}\xi^i\gamma_i\otimes \lambda_i,\; \exists B_{\delta}\subset u(M): \phi_i^t(B_{\delta})\subset u(A_{\epsilon}) \text{ for some } t \in (0,T)\},
            $$
where $u(M)$, $u(A_{\epsilon})$ and $\phi_i^t(B_{\delta})$ denote the images of the sets $M$, $A_{\epsilon}$ and $B_{\delta}$ under the functions $u$ and $\phi_i^t$.

\begin{center}
\includegraphics[scale=0.5]{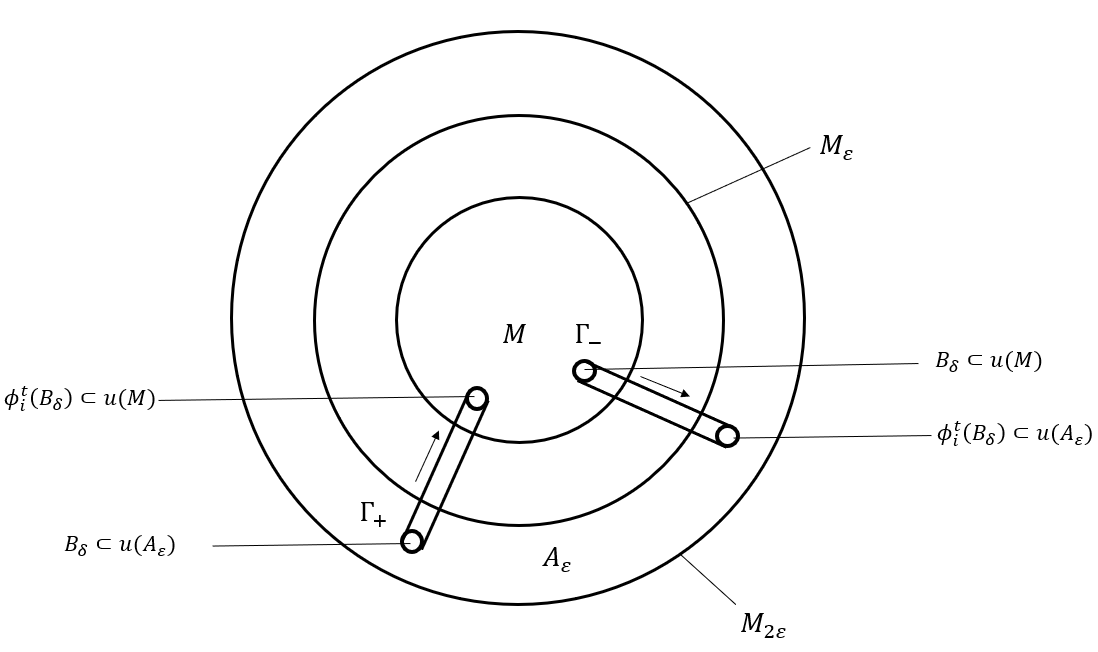} 
\end{center}
Fig 1 : The set $A_{\epsilon}$ and the region of superposition $M$ of two waves

\begin{defi}
\ \\
\noindent
The index $\Gamma_{M}$ of the region $M$ of superposition of two waves is defined as 
            \begin{equation}\label{indyk}
            \Gamma_{M}:= \card \Gamma_--\card \Gamma_+.
            \end{equation}
\end{defi}
\bigskip
Consequently, the index has the following properties
            \begin{itemize}
            \item[$\bullet$] $\Gamma_{M}\geqslant 0$\\ If $\Gamma_{M}=0$ then the number of waves entering and leaving the region of superposition is the same. $\Gamma_{M}>0$ means that additional waves result from the interaction.
            \item[$\bullet$] It was shown [3,7,8] that if the number of waves entering and leaving the interaction is the same (i.e. $\Gamma_{M}=0$) then the type of waves is also preserved.
            \item[$\bullet$] $\Gamma_{M}$ is independent of the choice of the functions $\xi^1,...,\xi^k$ (i.e. we have arbitrary wave profiles) and depends only on the vector fields $\gamma_1,...,\gamma_k$. 
            \item[$\bullet$]  $\Gamma_{M}$ is invariant with respect to diffeomorphisms of $M.$
            \end{itemize}

            \begin{defi}\label{elastic}
            \ \\
\noindent
            Superpositions of two single Riemann waves corresponding to $\gamma_1$ and $\gamma_2$ are called either
            \begin{itemize} 
            \item[$\bullet$]\textbf{elastic} if $\Gamma_{M}=0$\\
            or
            \item[$\bullet$]\textbf{non-elastic} if $\Gamma_{M}>0$.
            \end{itemize}
 
\end{defi}
\bigskip
Let us note that the notion of elastic superposition relates directly to the property of quasi-rectifiability of the vector fields $\{\gamma_1,...,\gamma_k\}$. Due to the fact that in the elastic case they satisfy the assumptions of Theorems 1.4, 1.5 and Corollary 1.6, these vector fields are quasi-rectifiable.

If the linearly independent vector fields $\{\gamma_1,...,\gamma_k\}$ commute among themselves, satisfying equation (\ref{rescvf}), and therefore forming a quasi-rectifiable family, then there exists a $k$-dimensional integral manifold $S$ on $\mathbb{R}^q$ parametrized in terms of $r^1,...,r^k$
\begin{equation}
\label{label17AC}
S:\quad u=\left(f^1\left(r^1,...,r^k\right),...,f^q\left(r^1,...,r^k\right)\right)\subset\mathbb{R}^q,
\end{equation}
obtained by integrating the system of partial differential equations (PDEs)
\begin{equation}
\label{label16AC}
\frac{\partial f^{\alpha}}{\partial r^s}=h_s\gamma_s^{\alpha}\left(f^1,...,f^q\right)\quad\alpha=1,...,q,\quad s=1,...,k
\end{equation}

It was shown [3] that if the set of implicit relations between variables $r$ and $x$
\begin{equation}\label{newequation1A}
\sum_{i=1}^2\lambda_i^s(r^1,...,r^k)x^i=\phi^s(r^1,...,r^k),\quad s=1,...,k
\end{equation}
for certain arbitrary functions $\phi^s$ can be solved (i.e. $r$ can be given as a graph over an open subset of $\mathbb{R}^2$), then the functions
\begin{equation}\label{newequation2}
u^\alpha=f^\alpha(r^1(x),...,r^k(x))
\end{equation}
constitute a Riemann k-wave solution of the initial system. 
The functions $r^1(x),...,r^k(x)$ are Riemann invariants. Let us note that if we assume all but two Riemann invariants in (\ref{label17AC}) and (\ref{newequation1A}) to be constants, then we obtain the solution (\ref{newequation2}) in the form of a double wave. Thus, a superposition of any two single waves, corresponding to any two vector fields $\gamma_i$ and $\gamma_j$, $i\neq j\in \{1,...,k\}$, is an elastic one.

In particular, we have

\begin{tw}$[6]$
\ \\
\noindent
The superposition of two single Riemann waves associated with $\gamma_1, \gamma_2,$ respectively, is elastic if and only if the family of vector fields $\{\gamma_1, \gamma_2\}$ is quasi-rectifiable.
\end{tw}




\section{Elastic wave superposition in the Euler system}

The one-dimensional Euler system describing the non-stationary compressible fluid flow is governed by the following system of three hyperbolic equations
\begin{equation}\label{euler}
				\begin{aligned}
                &{{\mathlarger{\frac{\partial}{\partial t}}}}\begin{pmatrix}
					\rho\\
					p\\
					u
				\end{pmatrix}=
				\begin{pmatrix}
					u&0&\rho\\
					0&u&\kappa p\\
					0&1/\rho&u
				\end{pmatrix}
				{\mathlarger{\frac{\partial}{\partial x}}}\begin{pmatrix}
					\rho\\
					p\\
					u
				\end{pmatrix},\\\quad 
                &v=(\rho,p,u)\in\mathbb{R}^3,\quad (x,t)\in\mathbb{R}^2,\quad \rho>0,\quad \kappa:=c_p/c_V>0.
			\end{aligned}
            \end{equation}
where $\rho$, $p$ and $u$ are the density, pressure and velocity of the fluid, respectively, and $\kappa$ is the adiabatic exponent.
The eigenvalues and eigenvectors for system \eqref{euler} correspond to three types of waves, two acoustic ones $S_+$ and $S_-$ and one entropic one $E$
\begin{equation}\label{eigenval}
    \begin{aligned}
        &S_+:\hspace{4mm}v_+=u+\sqrt{\frac{\kappa p}{\rho}},\quad \gamma_+ = (\rho,\kappa p,\sqrt{\frac{\kappa p}{\rho}}),\\
        &S_-:\hspace{4mm}v_-=u-\sqrt{\frac{\kappa p}{\rho}},\quad \gamma_-=(\rho,\kappa p,-\sqrt{\frac{\kappa p}{\rho}}),\\
               &E:\hspace{4mm}v_0=u,\quad\qquad\qquad \gamma_0=(1,0,0).
    \end{aligned}
\end{equation}
The sign $\pm$ means that the considered wave goes in the right or left direction with respect to the medium. Note that the vector fields $\{\gamma_+,\gamma_0,\gamma_-\}$ are linearly independent
			\[
			\gamma_+\wedge\gamma_0\wedge\gamma_-\neq 0\,\, \text{at any point } v\in \mathbb{R}^3.
			\]
Simple Riemann wave solutions of the Euler system (\ref{euler}) obtained in accordance with Theorem 1.1 have the form 
\begin{itemize}
    \item[$\bullet$] entropic wave $E$:
    $\rho$ changes arbitrarily and $p,u$ are constant
    \begin{equation*}
    \rho_t+u_0\rho_x=0\hspace{5mm}\Longleftrightarrow\hspace{5mm}\rho=\rho(x^2-u_0t),
    \end{equation*}    
    \item[$\bullet$] sound waves $S_{\pm}$:
    $p=A\rho^\kappa+p_0,$ $u = \frac{2}{\kappa-1}\sqrt{\kappa A\rho^{\frac{\kappa-1}{2}}}+u_0$ where $A$, $p_0$ and $u_0$ are arbitrary constants 
    \begin{equation*}
    \rho_t+\bigg{[}\dfrac{2}{\kappa-1}\sqrt{\kappa A\rho^{\frac{\kappa-1}{2}}}\pm\sqrt{\kappa A\rho^{\kappa-1}}\bigg{]}\rho_x=0.
    \end{equation*}
\end{itemize}
\bigskip

The superposition of single waves in the Euler system (\ref{euler}) was first investigated by B. Riemann ${[1,2]}$ in the case of two sound waves $S_+$ and $S_-$, for which he obtained a double wave solution using the method of characteristics. It was shown [4] that his result can be recreated using the approach presented here based on the commutator analysis.

The commutator relation for the vector fields $\gamma_+$ and $\gamma_-$ is given by
\begin{equation}
 \left[\gamma_+,\gamma_-\right]=\dfrac{1-\kappa}{2}\gamma_++\dfrac{\kappa-1}{2}\gamma_-.
 \label{extrathing1}
 \end{equation}
As the pair of vector fields $\{\gamma_+,\gamma_-\}$ is quasi-rectifiable, which corresponds to the elastic superposition of the sound waves $S_+$ and $S_-$, we can construct the rescaling functions $h_i=(\kappa p\rho^{-1})^{-1/2}$, $i=+,-$, such that
\begin{equation}
[h_+\gamma_+,h_-\gamma_-]=0.
\label{extrathing2}
\end{equation}
This fact allows us to parametrize the surface of the elastic superposition $v(M)$ in terms of the Riemann invariants $r^1$ and $r^2$. The double wave solution $\{S_+,S_-\}$ is given by
\begin{equation}
u = \kappa^{\frac{1}{2}}(r^1-r^2)+u_0,\quad \rho = Ae^{r^1+r^2},\quad p=\kappa Ae^{r^1+r^2}+p_0,
\label{extrathing3}
\end{equation}
where the invariants $r^1$ and $r^2$ satisfy the equations

\begin{equation}
\label{4.4}
\frac{\partial}{\partial t}\begin{pmatrix}
        r_1\\
        r_2
    \end{pmatrix}+
    \begin{pmatrix}
        \kappa^{\frac{1}{2}}(r^1-r^2+1)+u_0&0\\
        0&\kappa^{\frac{1}{2}}(r^1-r^2-1)+u_0
    \end{pmatrix}
    \frac{\partial}{\partial x}
    \begin{pmatrix}
        r_1\\
       	r_2
    \end{pmatrix}=\begin{pmatrix}
        0\\
       	0
    \end{pmatrix},
\end{equation}
which constitute a reduced form of the Euler system for this case.

It has been shown [3,7] that if the initial data for the system (\ref{4.4}) is sufficiently small and has compact and disjoint supports, then we can locally construct its solutions and consequently produce the double wave solutions of the Euler system (\ref{euler}).

The superposition of the sound waves $S_+$ and $S_-$ is the only type of elastic superposition admitted by the Euler system (\ref{euler}). Superpositions of two different types of simple waves, i.e. $S_+E$ and $S_-E$ are non-elastic due to the fact that the commutator relations for the corresponding pairs of vector fields are spanned by three vector fields, namely [4]
\begin{equation}\label{eulercom}
    \begin{aligned}
      &\left[\gamma_+,\gamma_0\right]=\dfrac{1}{4\rho}\gamma_+-\dfrac{1}{4\rho}\gamma_--\gamma_0,\\
      &\left[\gamma_-,\gamma_0\right]=-\dfrac{1}{4\rho}\gamma_++\dfrac{1}{4\rho}\gamma_--\gamma_0.
    \end{aligned}
\end{equation}
Consequently, the vector fields $\{\gamma_+,\gamma_0,\gamma_-\}$ do not constitute a quasi-rectifiable family of vector fields.

The next sections are devoted to presenting new tools for the analysis of non-elastic wave superpositions.


\section{Infinite-dimensional Lie algebra}

The family of vector fields corresponding to the Euler system (\ref{euler}) constitutes, by definition, a Lie module $\{|\gamma_+,\gamma_0,\gamma_-|\}$ since these vector fields are closed with respect to the Lie bracket (\ref{eulercom}) and (\ref{extrathing1}). In what follows we make the assumption that the $C^{\infty}$-Lie module $\{|\gamma_+,\gamma_0,\gamma_-|\}$ can be identified with an infinite-dimensional real Lie algebra. The following theorem describes the properties of this algebra.
    \begin{tw} $[6]$\label{infimp}
    \ \\
    \noindent
    The smallest (with respect to inclusion) real Lie algebra containing $\{\gamma_+,\; \gamma_-,\; \gamma_0\}$ is isomorphic to the algebras
    $$
    \mathcal{K} \simeq I^- \oplusrhrim_{\Theta} \boldsymbol{K}^-\qquad\mbox{or}\qquad \mathcal{H}\simeq I^+ \oplusrhrim_{\Theta} \boldsymbol{K}^+,
    $$
    where $I^-$ and $I^+$ are infinite-dimensional real Abelian Lie algebras, $\Theta$ is a shift operator, and $\boldsymbol{K}^\pm$ is a $3$-dimensional real Lie algebra which is a direct sum of the unique $2$-dimensional non-Abelian real Lie algebra and the unique $1$-dimensional real Lie algebra. The Abelian ideals are given by
    $$
    I^- = \spam \{\rho^{-1}(\gamma_+-\gamma_-),\rho^{-2}(\gamma_+-\gamma_-),\rho^{-3}(\gamma_+-\gamma_-),...\},
    $$
    $$
    I^+ = \spam \{\rho^{-1}(\gamma_++\gamma_-),\rho^{-2}(\gamma_++\gamma_-),\rho^{-3}(\gamma_++\gamma_-),...\}.
    $$
\end{tw}
Consequently, the decomposition $\mathcal{K} \simeq I^- \oplusrhrim_{\Theta} \boldsymbol{K}^-$ (and similarly $\mathcal{H}\simeq I^+ \oplusrhrim_{\Theta} \boldsymbol{K}^+$) leads to the following conclusions:

\bigskip

\begin{itemize}
    \item[$\bullet$] Due to the fact that $I^-\subset \mathcal{K}^-$ is an Abelian ideal, the whole qualitative behavior of wave interactions is encoded in $\boldsymbol{K}^-$.
    \item[$\bullet$] The quantitative aspect of the interactions involving the varying density $\rho$ is encoded in the infinite-dimensional component $I^-$.
    \item[$\bullet$] The infinite-dimensional character of the Lie algebra $\mathcal{K}$ comes only from the fact that the Lie algebra of $\mathcal{K}$ also contains `higher-order' iterations corresponding to sequences of wave superpositions.
\end{itemize}

As an interesting aside, beyond the immediate subject of this section, we can add two theorems concerning ``free algebras" containing multiplications of the vector fields $\{\gamma_+,\gamma_0,\gamma_-\}$ and $\{\gamma_+,\gamma_-\}$.

\begin{tw} $[6]$\label{witt_thm}
\ \\
\noindent
         The smallest real Lie algebra containing the vector fields $\{\rho^{-n}\gamma_+,\rho^{-n}\gamma_0,\rho^{-n}\gamma_-\},\; n\in \{0,1,2,3,...\}$ is isomorphic to the Virasoro algebra
        $$
        \mathcal{L} \simeq I_2 \oplusrhrim_{\Phi}(I_1 \oplusrhrim_{\Psi} \text{Witt})
        $$
        where $I_1,I_2$ are Abelian ideals, and $\Phi, \Psi$ are shift operators.
        \end{tw}
\begin{tw} $[6]$
\ \\
\noindent
The smallest (with respect to inclusion) real Lie algebra $\mathcal{B}$ containing the vector fields $\gamma_+$ and $\gamma_-$ and their consecutive multiplications
$
\{a_n=\rho^{-n}\gamma_+,\quad b_n=\rho^{-n}\gamma_-\},\quad n\in \{0,1,2,3,...\}
$
is isomorphic to
\begin{equation*}
\mathcal{B}\simeq I_3\oplusrhrim_\mu \text{Witt},
\end{equation*}
where $I_3$ is an Abelian Lie algebra and $\mu$ is a shift operator.
\end{tw}



\section{Parametrization of the region of wave superpositions}

The Lie algebra $\boldsymbol{K}^\pm$ describing the qualitative behavior of wave interactions can be obtained by transforming the initially given Lie module $\{|\gamma_-,\gamma_0,\gamma_+|\}$.
\bigskip

\begin{tw} $[6]$
\ \\
\noindent
The Lie module $\{|\gamma_+,\gamma_0,\gamma_-|\}$ corresponding to the Euler system (\ref{euler}) can be transformed by an angle preserving transformation into the unique (up to isomorphism) real Lie algebra isomorphic to the Lie algebra $\boldsymbol{K}^\pm$. 
\end{tw}

In order to determine the parametrization of the region of wave superpositions corresponding to the Lie module $\{|\gamma_+,\gamma_0,\gamma_-|\}$, we require that the algebra $\boldsymbol{K}^-$ (and similarly $\boldsymbol{K}^+$) be quasi-rectifiable. To this end, we rescale the algebra $\boldsymbol{K}^-$ through the introduction of a new basis of vector fields such that
\begin{equation}
w_1:=\gamma_++\gamma_-,\quad w_2:= \gamma_+-\gamma_-. 
\label{sec5extra1}
\end{equation}
Then the family of vector fields $\{w_1, w_2, \gamma_0\}$ is quasi-rectifiable. We determine the rescaling functions for the vector fields $\{w_1, w_2, \gamma_0\}$ using Theorem 1.4 and Corollary 1.6. We then construct the manifold $v(M)$ corresponding to the non-elastic wave superposition (for the case when the adiabatic exponent $\kappa=3$)
\begin{equation}\label{para}
v(M):\quad v=(\rho,p,u)={f(t_1,t_2,t_3)=(e^{2t_1+t_3},e^{6t_1},2\sqrt{3}t_2)},
\end{equation}
parametrized by certain functions $t_1$, $t_2$, $t_3$, where
\[
\frac{\partial(\rho,p,u)}{\partial (t_1,t_2,t_3)}\neq 0.
\]

Let us describe the manifold $v(M)$ in terms of the flow of the surfaces spanned by the vector fields $\{\gamma_+,\gamma_-\}$ along the vector field $\gamma_0$. 

Let $S_1$ be a manifold defined in a parametric form by (\ref{para}) and $S_2=\ln{S_1}$ be a manifold expressed in parametric form by $\beta(t_1,t_2,t_3)=\ln f(t_1,t_2,t_3)$. The foliations of the manifolds $S_1$ and $S_2$ are slicing them into stacks of quasi-rectifiable surfaces (leaves) denoted by $\Phi(t_3)$ and $\Sigma(t_3)$, respectively, which take the form
\begin{equation}\label{sigma-phi}
\begin{aligned}
&\Phi(t_3) =\spam\{\gamma_+,\gamma_-\}=\exp\Sigma(t_3) =\{ (e^{2t_1}e^{t_3},e^{6t_1},2\sqrt{3}t_2): t_1,t_2>0\}\subset \mathbb{R}^3,\\
&\Sigma(t_3)=\spam\{\frac{\partial \beta}{\partial t_1},\frac{\partial \beta}{\partial t_2}\} = \{(2t_1 + t_3, 6 t_1, \ln2\sqrt{3} + \ln t_2): t_1,t_2>0\}\subset \mathbb{R}^3,
\end{aligned}
\end{equation}
\begin{tw} $[6]$
\ \\
\noindent
The manifolds $S_1$ and $S_2$ corresponding to the non-elastic superpositions of waves related to the vector fields $\{\gamma_-,\gamma_0,\gamma_+\}$ can be obtained from the parallel transport of the quasi-rectifiable surfaces $\Phi(t_3)$ and $\Sigma(t_3)$ for $0<t_3<t'_3$, i.e.
\begin{equation}\label{splitty}
\begin{split}
S_1 = \bigcup_{t_3>0}\Phi(t_3),\quad & S_2 = \bigcup_{t_3>0}\Sigma(t_3),
\\
\Phi(t_3)\cap \Phi(t_3')=\emptyset,\qquad & \Sigma(t_3)\cap \Sigma(t_3')=\emptyset.
\end{split}
\end{equation}
The evolution of $\Phi(t_3),\; t_3>0$ can be completely reconstructed from the evolution of $\Sigma(t_3),\; t_3>0.$ 
\end{tw}

\begin{center}
\includegraphics[scale=0.7]{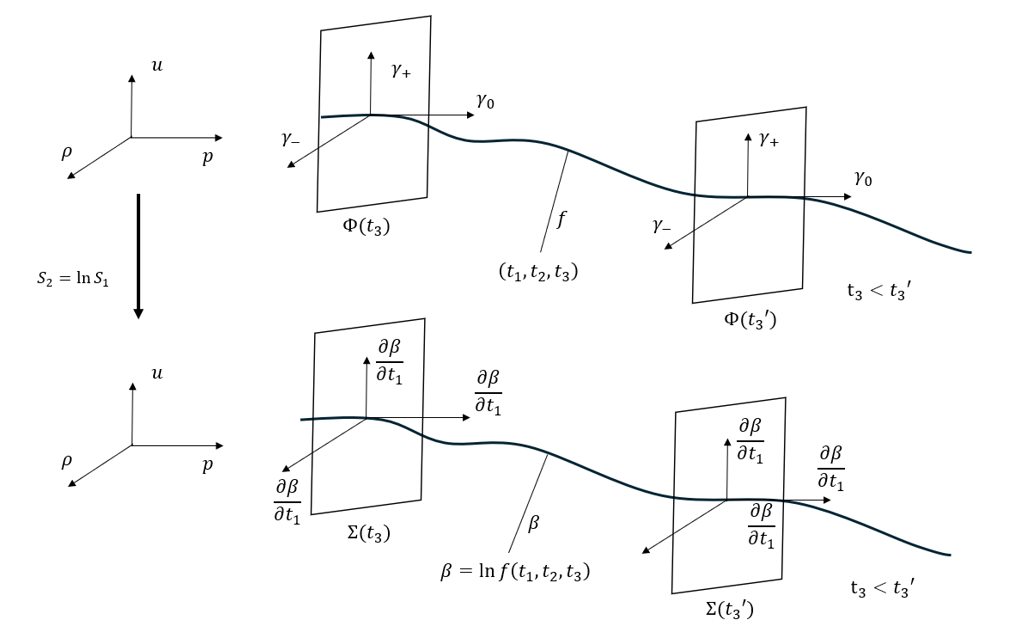} 
\end{center}
Fig 2 : Evolution of surfaces $\Phi(t_3)$ and $\Sigma(t_3)$.
\vspace{9mm}

\noindent The second fundamental form $II$ of the surface $\Phi(t_3)$ is  
\begin{equation}
\begin{aligned}
    &II = \begin{pmatrix}
        L & M\\
        M & N
    \end{pmatrix}=
    \begin{pmatrix}
        L&0\\
        0&0
    \end{pmatrix},\quad 
    \text{ where }\quad L=\frac{48 e^{t_3}e^{6t_1}}{(9e^{8t_1}+e^{2t_3})^{\frac{1}{2}}}.
\end{aligned}
\end{equation}
\bigskip

\noindent The Gaussian curvature of the surface $\Phi(t_3)$ is given by $K=k_1k_2 = 0,$ and the mean curvature is $H=\frac{1}{2}(k_1+k_2)=\frac{L}{2}.$

\section{Reduced form of the Euler system}

Using the parametrization (\ref{para}) of the manifold $S_1$  corresponding to non-elastic wave superpositions, we obtain the reduced form of the Euler system (\ref{euler}) (for $\kappa=3$) for
three dependent variables $(t_1,t_2,t_3)\in\mathbb{R}^3$ 

\begin{equation}
\label{10.1}
   \frac{\partial}{\partial t}\begin{pmatrix}
        t_1\\
        t_2\\
        t_3
    \end{pmatrix}=
    \begin{pmatrix}
        \frac{1}{2}&\frac{1}{2}&0\\
        \frac{1}{2h_2}&-\frac{1}{2h_2}&0\\
        0&0&\frac{1}{h_0}
    \end{pmatrix}
    \begin{pmatrix}
        v_+&0&0\\
        0&v_-&0\\
        0&0&v_0
    \end{pmatrix}
    \begin{pmatrix}
        1&h_2&0\\
        1&-h_2&0\\
        0&0&h_0
    \end{pmatrix}
    \frac{\partial}{\partial x}
    \begin{pmatrix}
        t_1\\
       	t_2\\
        t_3
    \end{pmatrix}.
\end{equation}
where $h_2(f)=\left(\dfrac{\rho}{p}\right)^{1/2}$ and $h_0(f)=\rho$ are rescaling functions for the vector fields $(\gamma_+-\gamma_-)$ and $\gamma_0$ respectively. 
Taking into account the eigenvalues (\ref{eigenval}) and the parametrization (\ref{para}), we obtain

\begin{equation}\label{matrix_mult}
\frac{\partial}{\partial t}\begin{pmatrix}
        t_1\\
        t_2\\
        t_3
    \end{pmatrix}=
    2\sqrt{3}\begin{pmatrix}
        t_2&\frac{1}{2}&0\\
        \frac{1}{2}e^{4t_1-t_3}&t_2&0\\
        0&0&t_2
    \end{pmatrix}
    \frac{\partial}{\partial x}
    \begin{pmatrix}
        t_1\\
       	t_2\\
        t_3
    \end{pmatrix}.
\end{equation}
System (\ref{matrix_mult}) constitutes the reduced form of the Euler system (\ref{euler}) and is an analogue of the reduced system (\ref{4.4}) for the elastic case. Note that the matrix in (\ref{matrix_mult}) has the Jordan form and is not diagonalizable as in the case of (\ref{4.4}). Systems of this form have been investigated (see e.g. [9], section 12.2) and it was shown that certain classes of their solutions can be obtained by the introduction of additional differential constraints of the first order. This subject will be addressed in a future work.

 Solutions of the system (\ref{matrix_mult}) describe the process of superpositions of an acoustic wave $S_+$ or $S_-$ and an entropic wave, which results in the production of a third wave in the region of interaction. This phenomenon was observed experimentally in wave-particle interactions in plasma physics [10,11,12].

\section{Surfaces with Lie group structure}

The simplified form (\ref{matrix_mult}) of the Euler system facilitates the analysis of certain geometric properties of the surfaces spanned by a pair of vector fields corresponding to interacting waves. We show that these surfaces have a Lie group structure. To this end we make use of the following three theorems.


\begin{prop}$[6]$
\ \\

    Let $\{X_1,X_2,X_3\}$ be a real Lie algebra and let $S$ be a surface spanned by the vector fields $X_1$ and $X_2$ in the neighbourhood of the point $p \in S.$ Let $\{X_1,X_2\}$ be a Lie subalgebra. Then locally, in the neighbourhood of $p,$ the surface $S$ has a Lie group structure corresponding to the algebra $\{X_1,X_2\}.$
\end{prop}

The next theorem defines the affine connection which plays an important role in our application. 
\bigskip

\begin{tw}[Nomizu ${[13]}$]
\ \\

    Let $G$ be a simply connected Lie group and let $\mathfrak{g}$ be the corresponding Lie algebra. Let the connection be given by
    \begin{equation}
    t(Y)(X) = \frac{1}{2}[X,Y]
    \end{equation}

     for any $X,Y \in \mathfrak{g}.$ Then the connection $t$ is a well-defined affine connection on the Lie group $G.$ Moreover, $t$ is both a left- and a right-invariant connection, and is the unique torsion-free connection where the $1$-parameter subgroups are geodesics.
\end{tw}
\bigskip

    The affine connection $t(Y)(X) = \frac{1}{2}[X,Y]$ does not need to be a Levi-Civita connection on the Lie group $G.$ This is because, in the general case, a bi-invariant metric on the Lie group $G$ does not exist.

Within these notions, we present a generalized counterpart of the result of section 5, which describes wave superpositions through deformations of quasi-rectifiable surfaces spanned by the vector fields $\{X,Y\}$.

\bigskip

\begin{tw}$[6]$
\ \\

    Let the real Lie algebra of vector fields $\{X,Y,Z\}$ take the form $$\{X,Y,Z\}=\{X,Y\}\oplus\{Z\}$$ and let $M$ be the simply-connected Lie group corresponding to $\{X,Y,Z\}.$ Then the Lie group $M$ is locally given through a parallel transport of the surfaces spanned by $\{X,Y\}$ and by the flow $\phi_Z$ of the vector field $Z$ in the affine connection $$t(A)(B)=\frac{1}{2}[A,B],\quad A,B \in \{X,Y,Z\}.$$
\end{tw}

Applying the above to the Euler system (\ref{euler}) we have

\begin{tw}$[6]$
\ \\

Let $\{|\gamma_+,\gamma_-,\gamma_0|\}$ be a Lie module corresponding to the Euler system (\ref{euler}). Let $\Phi$ be a surface spanned by the vector fields $\gamma_+$ and $\gamma_-$ on the neighbourhood of a point $p\in\Phi$; then, locally, in a neighbourhood of $p$, the surface $\Phi$ has a Lie group structure corresponding to the Lie algebra $\{\gamma_+,\gamma_-\}$.
\end{tw}

\begin{tw}$[6]$
\ \\

Let a Lie algebra of vector fields $\{w_1,w_2,\gamma_0\}$ corresponding to the Euler system (\ref{euler}), which has the form $$\{w_1,w_2,\gamma_0\}=\{w_1,w_2\}\oplus\{\gamma_0\}.$$ Let $\Phi$ be a simply connected Lie group corresponding to the algebra $\{w_1,w_2,\gamma_0\}$. Then the Lie group $\Phi$ is locally given by a parallel transport of the surfaces spanned by $\{w_1,w_2\}$ and by the flow of the vector field $\gamma_0$ in the unique torsion-free affine connection $$t(A)(B)=\frac{1}{2}[A,B],\text{ for any }A,B\in\{w_1,w_2\}$$ where the one-parameter subgroups of $\Phi$ are geodesics.
\end{tw}

\subsection*{Acknowledgements}

AMG has been supported by a research grant from NSERC of Canada. Both authors thank the Laboratoire de Physique Math\'ematique, Centre de Recherches Math\'ematiques, Universit\'e de Montr\'eal for its financial support.

\label{lastpage}
\end{document}